\documentclass[12pt,3p,times,authoryear]{elsarticle}

\usepackage{graphicx}

\usepackage[hyphenbreaks]{breakurl}
\usepackage[hyphens]{url}

\usepackage{smartdiagram}
\smartdiagramset{set color list={
		gray!30,
		gray!30,
		gray!30,
		gray!30,
		gray!30,
		gray!30,
	},
	sequence item border color=black,
	sequence item font size=16,
}

\usepackage{float}

\usepackage{mathptmx}      
\usepackage{latexsym}

\usepackage{float}

\usepackage{color}
\usepackage{colortbl}

\usepackage{amssymb}
\usepackage{comment}
\usepackage[multiple]{footmisc}
\usepackage{amsmath}

\usepackage{lipsum}
\usepackage{pfnote}
\usepackage{multirow}
\usepackage{float}
\usepackage{pgfplots}
\usepackage{tikz}
\usepackage{xcolor}
\usetikzlibrary{patterns}
\usepackage{amstext}

\usepackage{color,soul}

\usepackage{colortbl}

\usepackage{amsmath}

\usepackage{enumitem}

\usepackage{colortbl}

\definecolor{maroon}{cmyk}{0,0.87,0.68,0.32}

\usetikzlibrary{positioning,calc}

\usepackage{tikz}
\usetikzlibrary{trees}

\usetikzlibrary{positioning,decorations.text}

\tikzset{level 1/.style={level distance=2cm, sibling distance=10cm}}
\tikzset{level 2/.style={level distance=2cm, sibling distance=3cm}}

\tikzset{bag/.style={text width=10em, text centered,yshift=-0.5cm}}

\usetikzlibrary{positioning,calc}

\def\addlegendimage{\csname pgfplots@addlegendimage\endcsname}

\begin{document}

	\begin{frontmatter}
		
		
		
		\title{Characterizing Diabetes, Diet, Exercise, and Obesity Comments on Twitter}
		
		\author[label1]{Amir Karami}
		\author[label2]{Alicia A. Dahl}
		\author[label3]{Gabrielle Turner-McGrievy}
		\author[label4]{Hadi Kharrazi}
		\author[label5]{Jr. George Shaw}
		\address[label1]{University of South Carolina, School of Library and Information Science, \textbf{Email}: karami@sc.edu}
		\address[label2]{University of South Carolina, Arnold School of Public Health, \textbf{Email}: adahl@email.sc.edu}
		\address[label3]{University of South Carolina, Arnold School of Public Health, \textbf{Email}: mcgrievy@mailbox.sc.edu}
		\address[label4]{Johns Hopkins University, Bloomberg School of Public Health, \textbf{Email}: kharrazi@jhu.edu}
		\address[label5]{University of South Carolina, School of Library and Information Science, \textbf{Email}: gshaw@email.sc.edu}

		



\begin{abstract}
Social media provide a platform for users to express their opinions and share information. Understanding public health opinions on social media, such as Twitter, offers a unique approach to characterizing common health issues such as diabetes, diet, exercise, and obesity (DDEO); however, collecting and analyzing a large scale conversational public health data set is a challenging research task. The goal of this research is to analyze the characteristics of the general public's opinions in regard to diabetes, diet, exercise and obesity (DDEO) as expressed on Twitter. A multi-component semantic and linguistic framework was developed to collect Twitter data, discover topics of interest about DDEO, and analyze the topics. From the extracted 4.5 million tweets, 8\% of tweets discussed diabetes, 23.7\% diet, 16.6\% exercise, and 51.7\% obesity. The strongest correlation among the topics was determined between exercise and obesity ($p<.0002$). Other notable correlations were: diabetes and obesity ($p<.0005$), and diet and obesity ($p<.001$). DDEO terms were also identified as subtopics of each of the DDEO topics. The frequent subtopics discussed along with ``Diabetes", excluding the DDEO terms themselves, were blood pressure, heart attack, yoga, and Alzheimer. The non-DDEO subtopics for ``Diet" included vegetarian, pregnancy, celebrities, weight loss, religious, and mental health, while subtopics for ``Exercise" included computer games, brain, fitness, and daily plan. Non-DDEO subtopics for ``Obesity" included Alzheimer, cancer, and children. With 2.67 billion social media users in 2016, publicly available data such as Twitter posts can be utilized to support clinical providers, public health experts, and social scientists in better understanding common public opinions in regard to diabetes, diet, exercise, and obesity.

\end{abstract}

	\end{frontmatter}

	\section{Introduction}
	\label{Int}

The global prevalence of obesity has doubled between 1980 and 2014, with more than 1.9 billion adults considered as overweight and over 600 million adults considered as obese in 2014 \citep{who}. Since the 1970s, obesity has risen 37 percent affecting 25 percent of the U.S. adults \citep{flegal2012prevalence}. Similar upward trends of obesity have been found in youth populations, with a 60\% increase in preschool aged children between 1990 and 2010 \citep{harvardhsph}. Overweight and obesity are the fifth leading risk for global deaths according to the European Association for the Study of Obesity \citep{who}. Excess energy intake and inadequate energy expenditure both contribute to weight gain and diabetes \citep{hill2012energy,wing2001behavioral}.

Obesity can be reduced through modifiable lifestyle behaviors such as diet and exercise \citep{wing2001behavioral}. There are several comorbidities associated with being overweight or obese, such as diabetes \citep{kopelman2000obesity}. The prevalence of diabetes in adults has risen globally from 4.7\% in 1980 to 8.5\% in 2014. Current projections estimate that by 2050, 29 million Americans will be diagnosed with type 2 diabetes, which is a 165\% increase from the 11 million diagnosed in 2002 \citep{boyle2001projection}. Studies show that there are strong relations among diabetes, diet, exercise, and obesity (DDEO) \citep{hartz1983relationship,wing2001behavioral,barnard2009low,american2004physical}; however, the general public's perception of DDEO remains limited to survey-based studies \citep{tompson2012obesity}.

The growth of social media has provided a research opportunity to track public behaviors, information, and opinions about common health issues.  It is estimated that the number of social media users will increase from 2.34 billion in 2016 to 2.95 billion in 2020 \citep{Statista}. Twitter has 316 million users worldwide \citep{techcrunch} providing a unique opportunity to understand users' opinions with respect to the most common health issues \citep{mejova2015twitter}. Publicly available Twitter posts have facilitated data collection and leveraged the research at the intersection of public health and data science; thus, informing the research community of major opinions and topics of interest among the general population \citep{nasukawa2003sentiment,wiebe2003recognizing,zabin2008social} that cannot otherwise be collected through traditional means of research (e.g., surveys, interviews, focus groups) \citep{eichstaedt2015psychological,LQ4DY6_2015}. Furthermore, analyzing Twitter data can help health organizations such as state health departments and large healthcare systems to provide health advice and track health opinions of their populations and provide effective health advice when needed \citep{mejova2015twitter}.

Among computational methods to analyze tweets, computational linguistics is a well-known developed approach to gain insight into a population, track health issues, and discover new knowledge \citep{paul2011you,paul2012model,harris2014communication,zhao2011comparing}. Twitter data has been used for a wide range of health and non-health related applications, such as stock market \citep{bollen2011twitter} and election analysis \citep{tumasjan2010predicting}. Some examples of Twitter data analysis for health-related topics include: flu \citep{ritterman2009using,szomszor2010swineflu,lampos2010flu,lampos2012nowcasting,lampos2010tracking,culotta2010towards}, mental health \citep{coppersmith2015adhd}, Ebola \citep{lazard2015detecting,odlum2015can}, Zika \citep{fu2016people}, medication use \citep{scanfeld2010dissemination,hanson2013exploration,buntain2015your}, diabetes \citep{harris2013peer}, and weight loss and obesity \citep{dahl2016integrating,ghosh2013we,vickey2013twitter,turner2015tweet,harris2014communication}.

The previous Twitter studies have dealt with extracting common topics of one health issue discussed by the users to better understand common themes; however, this study utilizes an innovative approach to computationally analyze unstructured health related text data exchanged via Twitter to characterize health opinions regarding four common health issues, including diabetes, diet, exercise and obesity (DDEO) on a population level. This study identifies the characteristics of the most common health opinions with respect to DDEO and discloses public perception of the relationship among diabetes, diet, exercise and obesity. These common public opinions/topics and perceptions can be used by providers and public health agencies to better understand the common opinions of their population denominators in regard to DDEO, and reflect upon those opinions accordingly.

\section{Methods}
	\label{ME}
	
Our approach uses semantic and linguistics analyses for disclosing health characteristics of opinions in tweets containing DDEO words. The present study included three phases: data collection, topic discovery, and topic-content analysis.

\subsection{Data Collection}
\begin{figure*}[ht]

\begin{center}

\includegraphics[scale=0.3]{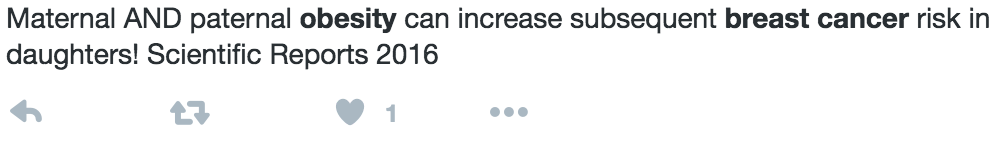} \\
\includegraphics[scale=0.3]{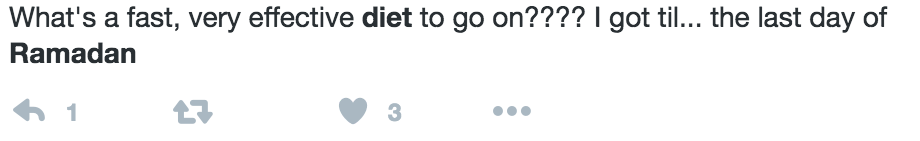} \\
\includegraphics[scale=0.3]{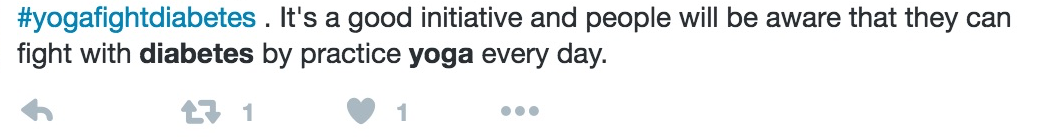} \\
\includegraphics[scale=0.3]{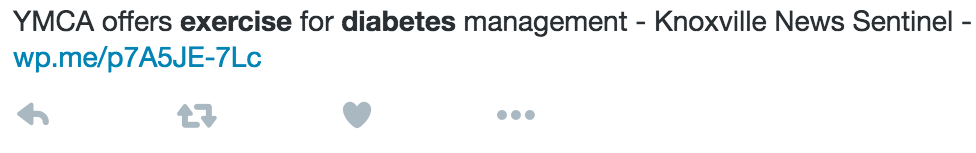}
\end{center}
\caption{A Sample of Tweets}
\label{fig:tweet_exp}
\end{figure*}

This phase collected tweets using Twitter's Application Programming Interfaces (API) \citep{twitter1}. Within the Twitter API, diabetes, diet, exercise, and obesity were selected as the related words \citep{wing2001behavioral} and the related health areas \citep{paul2011you}. Twitter's APIs provides both historic and real-time data collections. The latter method randomly collects 1\% of publicly available tweets. This paper used the real-time method to randomly collect 10\% of publicly available English tweets using several pre-defined DDEO-related queries (Table \ref{tab:query}) within a specific time frame. We used the queries to collect approximately 4.5 million related tweets between 06/01/2016 and 06/30/2016. The data will be available in the first author's website\footnote{\url{https://sites.google.com/site/karamihomepage/}}. Figure \ref{fig:tweet_exp} shows a sample of collected tweets in this research.

\subsection{Topic Discovery}
To discover topics from the collected tweets, we used a topic modeling approach that fuzzy clusters the semantically related words such as assigning ``diabetes", ``cancer", and ``influenza" into a topic that has an overall ``disease" theme \citep{karami2017fuzzy,karami2015fuzzy}. Topic modeling has a wide range of applications in health and medical domains such as predicting protein-protein relationships based on the literature knowledge \citep{asou2008predicting}, discovering relevant clinical concepts and structures in patients' health records~\citep{arnold2010clinical}, and identifying patterns of clinical events in a cohort of brain cancer patients~\citep{arnold2012topic}.

Among topic models, Latent Dirichlet Allocation (LDA) \citep{blei2003latent} is the most popular effective model \citep{lu2011investigating,paul2011you} as studies have shown that LDA is an effective computational linguistics model for discovering topics in a corpus  \citep{mcauliffe2008supervised,hong2010empirical}. LDA assumes that a corpus contains topics such that each word in each document can be assigned to the topics with different degrees of membership \citep{karami2015flatm,karami2015fuzzyiconf,karami2014fftm}. 

Twitter users can post their opinions or share information about a subject to the public. Identifying the main topics of users' tweets provides an interesting point of reference, but conceptualizing larger subtopics of millions of tweets can reveal valuable insight to users' opinions. The topic discovery component of the study approach uses LDA to find main topics, themes, and opinions in the collected tweets. 

We used the Mallet implementation of LDA \citep{blei2003latent,mccallum2002mallet} with its default settings to explore opinions in the tweets. Before identifying the opinions, two pre-processing steps were implemented: (1) using a standard list for removing stop words, that do not have semantic value for analysis (such as ``the"); and, (2) finding the optimum number of topics. To determine a proper number of topics, log-likelihood estimation with 80\% of tweets for training and 20\% of tweets for testing was used to find the highest log-likelihood, as it is the optimum number of topics \citep{wallach2009evaluation}. The highest log-likelihood was determined 425 topics.

\subsection{Topic Content Analysis}			
The topic content analysis component used an objective interpretation approach with a lexicon-based approach to analyze the content of topics. The lexicon-based approach uses dictionaries to disclose the semantic orientation of words in a topic. Linguistic Inquiry and Word Count (LIWC) is a linguistics analysis tool that reveals thoughts, feelings, personality, and motivations in a corpus \citep{karami2015online,karami2014exploiting,karami2014improving}. LIWC has accepted rate of sensitivity, specificity, and English proficiency measures \citep{golder2011diurnal}. LIWC has a health related dictionary that can help to find whether a topic contains words associated with health. In this analysis, we used LIWC to find health related topics.

\section{Results}
\label{Ex}

Obesity and Diabetes showed the highest and the lowest number of tweets (51.7\% and 8.0\%). Diet and Exercise formed 23.7\% and 16.6\% of the tweets (Table \ref{tab:query}).

\begin{table*}[ht]
	
	\caption{DDEO Queries}
	\begin{center}
		\begin{tabular}{  p{2.4cm}  p{6.5 cm} p{3.4cm} p{2.5cm}} 
			\hline
			{\cellcolor[gray]{.9}}  \textbf{Health Issue} & {\cellcolor[gray]{.9}}  \textbf{Queries} & {\cellcolor[gray]{.9}}  \textbf{Number of Tweets} & {\cellcolor[gray]{.9}}  \textbf{Percentage}  \\  		\hline
			Diabetes & diabetes OR \#diabetes  &   353,655 & 8.0\%  \\  			\hline
			Diet & diet OR \#diet OR dieting & 1,045,374 & 23.7\% \\ 			\hline 
			Exercise & exercise OR \#exercise OR exercising & 734,118 & 16.6\% \\  			\hline
			Obesity & obesity OR \#obesity OR fat  & 2,283,517 & 51.7\%  \\  			\hline

		\end{tabular}
		
		\label{tab:query}
	\end{center} 
\end{table*}

Out of all 4.5 million DDEO-related tweets returned by Tweeter's API, the LDA found 425 topics. We used LIWC to filter the detected 425 topics and found 222 health-related topics. Additionally, we labeled topics based on the availability of DDEO words. For example, if a topic had ``diet", we labeled it as a diet-related topic. As expected and driven by the initial Tweeter API's query, common topics were Diabetes, Diet, Exercise, and Obesity (DDEO). (Table \ref{tab:subtop}) shows that the highest and the lowest number of topics were related to exercise and diabetes (80 and 21 out of 222). Diet and Obesity had almost similar rates (58 and 63 out of 222).

Each of the DDEO topics included several common subtopics including both DDEO and non-DDEO terms discovered by the LDA algorithm (Table \ref{tab:subtop}). Common subtopics for ``Diabetes", in order of frequency, included type 2 diabetes, obesity, diet, exercise, blood pressure, heart attack, yoga, and Alzheimer. Common subtopics for ``Diet" included obesity, exercise, weight loss [medicine], celebrities, vegetarian, diabetes,  religious diet, pregnancy, and mental health. Frequent subtopics for ``Exercise" included fitness, obesity, daily plan, diet, brain, diabetes, and computer games. And finally, the most common subtopics for ``Obesity" included diet, exercise, children, diabetes, Alzheimer, and cancer (Table \ref{tab:subtop}). Table \ref{tab:topexample} provides illustrative examples for each of the topics and subtopics.  

\begin{table*}[ht]
	\Large
	\caption{DDEO Topics and Subtopics - Diabetes, Diet, Exercise, and Obesity are shown with
		italic and underline styles in subtopics
	}
	\begin{center}
		\scalebox{0.5}{
			\begin{tabular}{  c c c c c c  c c } 
				\hline
				
				{\cellcolor[gray]{.9}}  \textbf{Topics} & {\cellcolor[gray]{.9}}  \textbf{Frequency} & 
				{\cellcolor[gray]{.9}} \textbf{Subtopics} & {\cellcolor[gray]{.9}}  \textbf{Distributions (\%)} & {\cellcolor[gray]{.9}}  \textbf{Topics} & {\cellcolor[gray]{.9}}  \textbf{Frequency} &
				{\cellcolor[gray]{.9}} \textbf{Subtopics} & {\cellcolor[gray]{.9}}  \textbf{Distributions (\%)}  \\ 
				\hline
				\textbf{Diabetes} & 21 & Diabetes Type 2 & 42.87\%  &  \textbf{Diet} & 63 &\underline{\textit{Obesity}}& 39.69\%  \\
				& & \underline{\textit{Obesity}} & 14.29\% & & & \underline{\textit{Exercise}}& 15.87\%  \\
				& & \underline{\textit{Diet}} & 9.52\% & & & Weight Loss & 12.71\% \\
				& & \underline{\textit{Exercise}}& 9.52\% & & & Celebrities & 9.52\%  \\
				& & Blood Pressure & 9.52\% & & & Vegetarian & 9.52\%\\
				& & Heart Attack& 4.76\%  & & & \underline{\textit{Diabetes}} & 3.17\% \\
				& & Yoga & 4.76\% & & &   Religious Diet & 3.17\%   \\
				& & Alzheimer & 4.76\% & & & Weight Loss Medicine & 3.17\%  \\
				& &	 &		&	& & Pregnancy & 1.59\% \\
				& &  &  &  & & Mental Health & 1.59\% \\

				\hline
				\textbf{Exercise} & 80  & Fitness & 32.5\%  & \textbf{Obesity} & 58  & \underline{\textit{Diet}} & 43.11\% \\
				& &  \underline{\textit{Obesity} } & 22.5\% & & & \underline{\textit{Exercise} } & 31.04\% \\
				& & Daily Plan & 21.25\%  & & & Children &  17.24\% \\ 
				& & \underline{\textit{Diet}} & 11.25\%   & & & \underline{\textit{Diabetes}} & 5.17\%  \\ 	
				& & Brain & 8.75\%  & & & Alzheimer  &  1.72\% \\ 	
				& &  \underline{\textit{Diabetes}}& 2.50\%   & & & Cancer & 1.72\% \\
				& &  Computer Games & 1.25\% & &\\
				
				\hline

			\end{tabular}}
			
			\label{tab:subtop}
		\end{center} 
	\end{table*}

Further exploration of the subtopics revealed additional patterns of interest (Tables \ref{tab:subtop} and \ref{tab:topexample}). We found 21 diabetes-related topics with 8 subtopics. While type 2 diabetes was the most frequent of the sub-topics, heart attack, Yoga, and Alzheimer are the least frequent subtopics for diabetes. Diet had a wide variety of emerging themes ranging from celebrity diet (e.g., Beyonce) to religious diet (e.g., Ramadan). Diet was detected in 63 topics with 10 subtopics; obesity, and pregnancy and mental health were the most and the least discussed obesity-related topics, respectively. Exploring the themes for Exercise subtopics revealed subjects such as computer games (e.g., Pokemon-Go) and brain exercises (e.g., memory improvement). Exercise had 7 subtopics with fitness as the most discussed subtopic and computer games as the least discussed subtopic. Finally, Obesity themes showed topics such as Alzheimer (e.g., research studies) and cancer (e.g., breast cancer). Obesity had the lowest diversity of subtopics: six with diet as the most discussed subtopic, and Alzheimer and cancer as the least discussed subtopics.

	\begin{table*}[ht]
		\centering			
		
		\Large	
		\caption{Topics Examples}
		\scalebox{0.4}{
			\begin{tabular}{ cccccccc}\hline
				\rowcolor[gray]{.9} \textbf{Blood Pressure}& \textbf{Heart Attack} & \textbf{Diabetes Type II}  & \textbf{Yoga} & \textbf{Alzheimer}  &  \textbf{Obesity}  & \textbf{Diet and Exercise} & \textbf{Obesity}   \\\hline
				
				risk      & heart           & change        & diabetes      & medicine          & diabetes  & helps & health    \\
				
				blood     & diabetes        & diabetes      & \#yogafightsdiabetes      & diseases          & surgery & diabetes & diet \\
				
				high     & cardiovascular   & \#lifestyle     & yoga      & common           & treatment  & children & obesity     \\
				
				diabetes     & attack           & type        &  control      & drugs            & obesity & exercise  & immune    \\
				
				pressure    & stroke            & ii           & life       & Alzheimer        & cure  & diet & syndrome \\ \hline

				\rowcolor[gray]{.9}	\textbf{Vegetarian} & \textbf{Pregnancy Diet} & \textbf{Celebrities Diet} & \textbf{Weight Loss Diet} & \textbf{Weight Loss Medicine}  & \textbf{Religious Diet} &   \textbf{Mental Health} &\textbf{Exercise\& Diabetes}   \\ \hline
				
				diet           & pregnancy       & diet           & weightlose      & diet &     burning & health  &  helps     \\
				
				eat            & motherhood      & beyonce        & effective     & \#weightloss &  \#weightloss  & nutrition & diabetes  \\
				
				fruits         & diet            & tips            & morning      & slimming   &  fasting & benefits & children      \\
				
				vegetables     & baby            & fatloss         & dieting      & pills  &   Ramadan  & healing & exercise       \\
				
				fresh         & motherhood      & \#angelinajolie   & banana      & \#fatburners &  diets & \#mentalhealth & diet     \\ \hline

				\rowcolor[gray]{.9} \textbf{Diet} & \textbf{Daily Plan} & \textbf{Computer Games} &    \textbf{Brain}  &  \textbf{Fitness} & \textbf{Diet\& Diabetes} &  \textbf{Obesity} & \textbf{Exercise}   \\\hline
				
				diet &food        & exercise              & exercise        & fitness  & helps & workout & bellyfat  \\
				
				exercise   & exercise    & finding            & brain              & \#gymlife  & diabetes & burning & losing \\
				
				protein    & calorie     & pokemon             & improve          & bodybuilding   & children & exercise & exercise\\
				
				beauty   &goal        & \#pokemongo            & memory        & gym & exercise & fatburn &   ways  \\
				
				muscle       &completed   & hour           & performance    & workout  & diet& obesity & effective \\ 
				
				\hline	
				
				\rowcolor[gray]{.9} \textbf{Diet} & \textbf{Alzheimer} & \textbf{Cancer} & \textbf{Children} & \textbf{Diabetes}      \\\hline
				
				health      & study        & cancer      & obesity & diabetes     \\
				
				diet         & link       & breast      & kids & surgery    \\
				
				obesity      & Alzheimer       & study         & childhood  & treatment \\
				
				immune       & obesity       & risk        &   rates & obesity   \\
				
				syndrome     & research      & obesity      & problem & cure   \\
				
				\hline

			\end{tabular}}

			\label{tab:topexample}

		\end{table*}

Diabetes subtopics show the relation between diabetes and exercise, diet, and obesity. Subtopics of diabetes revealed that users post about the relationship between diabetes and other diseases such as heart attack (Tables \ref{tab:subtop} and \ref{tab:topexample}).  The subtopic Alzheimer is also shown in the obesity subtopics. This overlap between categories prompts the discussion of research and linkages among obesity, diabetes, and Alzheimer's disease. Type 2 diabetes was another subtopic expressed by users and scientifically documented in the literature.

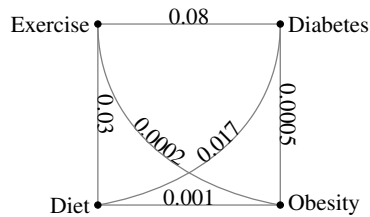
\begin{figure}[ht]
	\large
	\centering
	\scalebox{0.6}{
		\begin{tikzpicture}
		\draw[gray, thick] (0,0) -- (4,0);
		\draw[gray, thick] [decoration={text along path,
			text={0.08},text align={center}},decorate] (0,0) -- (4,0);
		
		\draw[gray, thick] [decoration={text along path,
			text={0.0002},text align={center}},decorate] (0,0) to [out=-90,in=170] (4,-4);
		\draw[gray, thick] (0,0) to [out=-90,in=170] (4,-4);
		
		\draw[gray, thick] (0,0) -- (0,-4);
		\draw[gray, thick] [decoration={text along path,
			text={0.03},text align={center}},decorate] (0,0) -- (0,-4);
		
		\draw[gray, thick] (4,0) -- (4,-4);
		\draw[gray, thick] [decoration={text along path,
			text={0.0005},text align={center}},decorate] (4,0) -- (4,-4);

		\draw[gray, thick] (0,-4) -- (4,-4);
		\draw[gray, thick] [decoration={text along path,
			text={0.001},text align={center}},decorate] (0,-4) -- (4,-4);

		\draw[gray, thick] (0,-4) to [out=10,in=-90] (4,0);
		\draw[gray, thick]  [decoration={text along path,
			text={0.017},text align={center}},decorate] (0,-4) to [out=10,in=-90] (4,0);

		\filldraw[black] (0,0) circle (2pt) node[anchor=east] {Exercise};
		\filldraw[black] (4,0) circle (2pt) node[anchor=west] {Diabetes };
		\filldraw[black] (0,-4) circle (2pt) node[anchor=east] {Diet};
		\filldraw[black] (4,-4) circle (2pt) node[anchor=west] {Obesity};
		
		\end{tikzpicture}}
	\caption{DDEO Correlation P-Value}
	\label{fig:areas}
\end{figure}

The main DDEO topics showed some level of interrelationship by appearing as subtopics of other DDEO topics. The words with \textit{italic} and \underline{underline} styles in Table 2 demonstrate the relation among the four DDEO areas. Our results show users' interest about posting their opinions, sharing information, and conversing about exercise \& diabetes, exercise \& diet, diabetes \& diet, diabetes \& obesity, and diet \& obesity (Figure \ref{fig:areas}). The strongest correlation among the topics was determined to be between exercise and obesity ($p<.0002$). Other notable correlations were: diabetes and obesity ($p<.0005$), and diet and obesity ($p<.001$).

\section{Discussion}
\label{Di}
Diabetes, diet, exercise, and obesity are common public health related opinions. Analyzing individual- level opinions by automated algorithmic techniques can be a useful approach to better characterize health opinions of a population. Traditional public health polls and surveys are limited by a small sample size; however, Twitter provides a platform to capture an array of opinions and shared information a expressed in the words of the tweeter. Studies show that Twitter data can be used to discover trending topics, and that there is a strong correlation between Twitter health conversations and Centers for Disease Control and Prevention (CDC) statistics \citep{prier2011identifying}. 

This research provides a computational content analysis approach to conduct a deep analysis using a large data set of  tweets. Our framework decodes public health opinions in DDEO related tweets, which can be applied to other public health issues. Among health-related subtopics, there are a wide range of topics from diseases to personal experiences such as participating in religious activities or vegetarian diets.

Diabetes subtopics showed the relationship between diabetes and exercise, diet, and obesity (Tables \ref{tab:subtop} and \ref{tab:topexample}). Subtopics of diabetes revealed that users posted about the relation between diabetes and other diseases such as heart attack. The subtopic Alzheimer is also shown in the obesity subtopics. This overlap between categories prompts the discussion of research and linkages among obesity, diabetes, and Alzheimer's disease. Type 2 diabetes was another subtopic that was also expressed by users and scientifically documented in the literature. The inclusion of Yoga in posts about diabetes is interesting. While yoga would certainly be labeled as a form of fitness, when considering the post, it was insightful to see discussion on the mental health benefits that yoga offers to those living with diabetes \citep{ross2010health}.

Diet had the highest number of subtopics. For example,  religious diet activities such as fasting during the month of Ramadan for Muslims incorporated two subtopics categorized under the diet topic (Tables \ref{tab:subtop} and \ref{tab:topexample}). This information has implications for the type of diets that are being practiced in the religious community, but may help inform religious scholars who focus on health and psychological conditions during fasting. Other religions such as Judaism, Christianity, and Taoism have periods of fasting that were not captured in our data collection, which may have been due to lack of posts or the timeframe in which we collected data. The diet plans of celebrities were also considered influential to explaining and informing diet opinions of Twitter users \citep{boyington2008cultural}.

Exercise themes show the Twitter users' association of exercise with ``brain" benefits such as increased memory and cognitive performance (Tables \ref{tab:subtop} and \ref{tab:topexample}) \citep{cotman2002exercise}. The topics also confirm that exercising is associated with controlling diabetes and assisting with meal planning \citep{laaksonen2005physical,american2004physical}, and obesity \citep{ross2000reduction}. Additionally, we found the Twitter users mentioned exercise topics about the use of computer games that assist with exercising. The recent mobile gaming phenomenon Pokeman-Go game \citep{pokemango} was highly associated with the exercise topic. Pokemon-Go allows users to operate in a virtual environment while simultaneously functioning in the real word. Capturing Pokemons, battling characters, and finding physical locations for meeting other users required physically activity to reach predefined locations. These themes reflect on the potential of augmented reality in increasing patients' physical activity levels \citep{schwarzer2008modeling}.

Obesity had the lowest number of subtopics in our study. Three of the subtopics were related to other diseases such as diabetes (Tables \ref{tab:subtop} and \ref{tab:topexample}). The scholarly literature has well documented the possible linkages between obesity and chronic diseases such as diabetes \citep{flegal2012prevalence} as supported by the study results. The topic of children is another prominent subtopic associated with obesity. There has been an increasing number of opinions in regard to child obesity and national health campaigns that have been developed to encourage physical activity among children \citep{nflplay60}. Alzheimer was also identified as a topic under obesity. Although considered a perplexing finding, recent studies have been conducted to identify possible correlation between obesity and Alzheimer's disease \citep{verdile2015inflammation,luchsinger2012central,kivipelto2005obesity}. Indeed, Twitter users have expressed opinions about the study of Alzheimer's disease and the linkage between these two topics.

This paper addresses a need for clinical providers, public health experts, and social scientists to utilize a large conversational dataset to collect and utilize population level opinions and information needs. Although our framework is applied to Twitter, the applications from this study can be used in patient communication devices monitored by physicians or weight management interventions with social media accounts, and support large scale population-wide initiatives to promote healthy behaviors and preventative measures for diabetes, diet, exercise, and obesity.

This research has some limitations. First, our DDEO analysis does not take geographical location of the Twitter users into consideration and thus does not reveal if certain geographical differences exists. Second, we used a limited number of queries to select the initial pool of tweets, thus perhaps missing tweets that may have been relevant to DDEO but have used unusual terms referenced. Third, our analysis only included tweets generated in one month; however, as our previous work has demonstrated \citep{turner2015tweet}, public opinion can change during a year. Additionally, we did not track individuals across time to detect changes in common themes discussed. Our future research plans includes introducing a dynamic framework to collect and analyze DDEO related tweets during extended time periods (multiple months) and incorporating spatial analysis of DDEO-related tweets.

\section{Conclusion}
		\label{Co}
This study represents the first step in developing routine processes to collect, analyze, and interpret DDEO-related posts to social media around health-related topics and presents a transdisciplinary approach to analyzing public discussions around health topics. With 2.34 billion social media users in 2016, the ability to collect and synthesize social media data will continue to grow. Developing methods to make this process more streamlined and robust will allow for more rapid identification of public health trends in real time. \\

\textbf{Note}: Amir Karami will handle correspondence at all stages of refereeing and publication.

\section{Conflict of interest}
The authors state that they have no conflict of interest.

\section{Acknowledgement}
This research was partially supported by the first author's startup research funding provided by the University of South Carolina, School of Library and Information Science. We thank Jill Chappell-Fail and Jeff Salter at the University of South Carolina College of Information and Communications for assistance with technical support. \\

\noindent \textbf{References} 

	\bibliographystyle{elsarticle-harv} 
	\bibliography{refrence}
	
	
		
		
		
\end{document}